%% file: paper.tex
\documentclass[a4paper]{article}

\usepackage{INTERSPEECH2022}

\title{Frequency Dynamic Convolution: Frequency-Adaptive Pattern Recognition for Sound Event Detection}
\name{Hyeonuk Nam, Seong-Hu Kim, Byeong-Yun Ko, Yong-Hwa Park}

\address{Department of Mechanical Engineering, Korea Advanced Institute of Science and Technology, Korea}
\email{\{frednam, seonghu.kim, b.y.ko, yhpark\}@kaist.ac.kr}

\begin{document}

\maketitle
\begin{abstract}
\input{sections/0_abstract}
\end{abstract}

\vspace{5pt}
\section{Introduction}
\vspace{5pt}
\input{sections/1_introduction}

\vspace{5pt}
\section{Proposed Method}
\vspace{5pt}
\input{sections/2_proposed_method}

\vspace{5pt}
\section{Experimental Setups}
\vspace{5pt}
\input{sections/3_experimental_setups}

\vspace{5pt}
\section{Results and Discussion}
\vspace{5pt}
\input{sections/4_results_and_discussion}

\vspace{5pt}
\section{Conclusions}
\vspace{5pt}
\input{sections/5_conclusion}

\vspace{5pt}
\section{Acknowledgements}
\vspace{5pt}
This work was supported by “Human Resources Program in Energy Technology” of the Korea Institute of Energy Technology Evaluation and Planning (KETEP), granted financial resource from the Ministry of Trade, Industry \& Energy, Republic of Korea. (No. 20204030200050), and also supported by Korea Institute of Marine Science and Technology Promotion(KIMST) grant funded by the year 2022 Finances of Korea Ministry of Oceans and Fisheries (MOF) (Development of Technology for Localization of Core Equipment in the Marine Fisheries Industry, 20210623).

\vfill\pagebreak

\bibliographystyle{IEEEtran}
\bibliography{reference.bib}

\end{document}

%% file: sections/0_abstract.tex
2D convolution is widely used in sound event detection (SED) to recognize two dimensional time-frequency patterns of sound events. However, 2D convolution enforces translation equivariance on sound events along both time and frequency axis while frequency is not shift-invariant dimension. In order to improve physical consistency of 2D convolution on SED, we propose \textit{frequency dynamic convolution} which applies kernel that adapts to frequency components of input. Frequency dynamic convolution outperforms the baseline by 6.3\% in DESED validation dataset in terms of polyphonic sound detection score (PSDS). It also significantly outperforms other pre-existing content-adaptive methods on SED. In addition, by comparing class-wise F1 scores of baseline and frequency dynamic convolution, we showed that frequency dynamic convolution is especially more effective for detection of non-stationary sound events with intricate time-frequency patterns. From this result, we verified that frequency dynamic convolution is superior in recognizing frequency-dependent patterns.

%% file: sections/1_introduction.tex
Sound event detection (SED), which aims to recognize sound event class and corresponding timestamps (onset and offset) from audio signals, has been rapidly growing since success of deep learning (DL) methods in various pattern recognition fields\cite{CASSE, sedmetrics, DCASEtask4, mytechreport, dcasewebsite}. SED has been adopting various DL methods from speech processing tasks such as automatic speech recognition (ASR) and speaker verification which are also based on audio signal processing \cite{specaug, conformer, coughcam, acnn}. However, there is no guarantee that DL methods from other domains are flawlessly compatible with SED. While transformer is prevalently used in natural language processing (NLP) and ASR \cite{transformer, bert}, it does not necessarily perform better than pre-existing convolutional recurrent neural network (CRNN) on SED \cite{dcase7wang,dcase22koo,dcase18hang}. Furthermore, conformer \cite{conformer} which achieved state-of-the-art performance in ASR failed to show stable performance in SED \cite{dcase21na, dcase3}. Considering the similarity between ASR and SED that both tasks take audio data as input and yield sequential output, conformer appears to be a reasonable choice for SED as well but it turned out not to be. This emphasizes that DL methods qualified in other similar domains have to be thoroughly examined before applying to SED.

\begin{figure}[ht]
\centerline{\includegraphics[width=8cm]{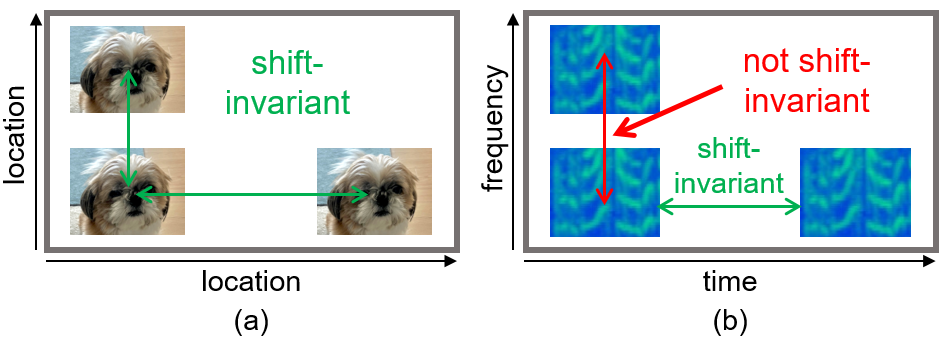}}
\vspace{-8pt}
\caption{An illustration of shift-invariance of (a) location in 2D image data, (b) time and frequency in 2D audio data (log Mel spectrogram).}
\label{fig:equivariance}
\vspace{-10pt}
\end{figure}

2D convolution has been widely used in speech and audio domain DL tasks to recognize 2D time-frequency patterns. However, 2D Convolutions is proposed to recognize 2D image data thus not exactly compatible with audio data. Recently, there have been several attempts to make 2D convolution to reflect domain knowledge of audio data rather than image data. For speaker recognition tasks, inspired by domain knowledge on speech rapidly changing along time due to variation of phonemes, temporally adaptive 2D convolution methods are proposed \cite{acnn, tdycnn, DTDY}.

Similarly, we apply domain knowledge on SED that sound events exhibit frequency-dependent patterns. On SED, 2D convolution imposes translation equivariance along both time and frequency axis. It is because 2D convolution is designed for 2D image data which is shift-invariant on both dimensions as illustrated in Figure 1 (a). However, translation equivariance should not be imposed along frequency axis because 2D audio data is not shift-invariant along frequency axis as shown in Figure 1 (b). A time-frequency pattern would sound different when it is shifted along the frequency axis. Therefore, it would be more physically consistent to release translation equivariance along frequency axis from 2D convolution in order to account for frequency-dependence of sound events. In addition, SED was shown to be highly frequency-dependent in previous work, FilterAugment \cite{filtaug}. Therefore, instead of simply applying pre-existing content-adaptive methods such as dynamic convolution using input-adaptive kernel \cite{dyconv} and temporal dynamic convolution using time-adaptive kernel \cite{tdycnn}, we propose a domain knowledge inspired method named \textit{frequency dynamic convolution}. The main contributions of this work are as follows:
\begin{enumerate}
\vspace{3pt}
    \itemsep0.3em
    \item{We propose \textit{frequency dynamic convolution} that applies frequency-adaptive kernel in order to release translation equivariance of 2D convolution along frequency axis, for physical consistency with the time-frequency patterns in sound events.}
    \item{Proposed frequency dynamic convolution outperforms not only baseline (by 6.3\%), but also other pre-existing content-adaptive methods proposed for other tasks (dynamic convolution and temporal dynamic convolution).}
    \item{By class-wise performance comparison with the baseline, we showed that frequency dynamic convolution is especially effective on non-stationary sound events, proving that frequency dynamic convolution is superior on frequency-dependent pattern recognition.}
\vspace{5pt}
\end{enumerate}
The official implementation code is available on GitHub\footnote{https://github.com/frednam93/FDY-SED}.

%% file: sections/2_proposed_method.tex
\subsection{Motivation}
\vspace{-5pt}
Dynamic convolution was proposed to enhance representation capability of vanilla convolution by adapting convolution kernel to given input \cite{dyconv}. Dynamic convolution first extracts attention weights from input of convolution, and then performs weighted sum of basis kernels with attention weights to obtain kernel optimal for the input. Similarly, temporal dynamic convolution is proposed for speaker verification to apply kernel optimal for each time frame. It uses kernel that adapts to input’s time frame in order to consider various phonemes composing spoken speech along time axis \cite{tdycnn}.

Likewise, we were motivated to make use of the domain knowledge on SED that sound event patterns are frequency-dependent: the same time-frequency pattern sounds different on different frequency regions. Within time-frequency domain, certain pattern sounds the same when it is shifted along time axis because it has the same frequency components despite it happen at different time. On the other hand, it would sound different when it is shifted along frequency axis because the frequency component which composes acoustic characteristics of the sound event changes. Such characteristics of time-frequency pattern is shown in Figure 1 (b). Small shift in frequency axis could be perceived as just minor pitch shift, but large shift would make it difficult for us to recognize the information of original sound pattern. In addition, previous work proved that frequency-dependency is a critical issue in SED. FilterAugment, a data augmentation method which applies different weights on random frequency regions proved that regularizing SED model over wider frequency regions enhances SED performance by a large margin, 6.5\% in terms of polyphonic sound detection score (PSDS) \cite{filtaug, PSDS}. This is because sound events exhibit distinctive patterns over various frequency regions. Such insights inspired us to develop a frequency-dependent pattern recognition method for SED.

Vast majority of SED models based on CRNN architecture \cite{crnn} uses 2D convolution which enforces translation equivariance on two dimensions of input. Computer vision domain utilizes translation equivariance of 2D convolution to recognize image pattern regardless of its relative position within the image \cite{physicsinformedML, vit}. Likewise, DL based audio tasks use 2D convolution on time-frequency patterns enforcing translation equivariance along both time and frequency axis. While translation equivariance is helpful on SED along time axis, it might not along frequency axis. 2D convolution introduces inconsistency between its frequency equivariance and the frequency-dependence of sound events. Thus we should maintain translation equivariance of 2D convolution on time dimension while loosening it on frequency dimension to improve model's physical consistency with sound events' time-frequency patterns and to improve SED performance. \textit{Frequency dynamic convolution} is proposed to solve this problem using dynamic kernel that adapts to the input’s frequency component.

\subsection{Frequency Dynamic Convolution}
\vspace{-5pt}
Frequency dynamic convolution uses frequency-adaptive kernel in order to enforce frequency-dependency on 2D convolution thus to improve physical consistency of SED model with sound events' time-frequency patterns. The operation is illustrated in Figure 2. It first extracts frequency-adaptive attention weights from input by applying average pooling over time axis followed by two 1D convolution layers along channel axis. Instead of using fully-connected (FC) layers as dynamic convolution did \cite{dyconv}, we applied 1D convolution in order to consider adjacent frequency components as well. Between two 1D convolution layers, batch normalization and ReLU are applied. 1D convolution layers compress channel dimension into the number of basis kernels. Then, softmax is applied to make frequency-adaptive attention weights range between zero and one and make sum of the weights for different basis kernel equal to one. Temperature of 31 was applied on the softmax to ensure uniform learning of basis kernels and stable training \cite{tdycnn, dyconv}. Then frequency-adaptive convolution kernel is obtained by weighted sum of basis kernels using frequency-adaptive attention weights, where basis kernels are trainable parameters as well. Obtained frequency-adaptive kernel is used for frequency dynamic convolution operation just as normal 2D convolution.

Note that the procedure described in the preceding paragraph is for better explanation and presentation of the concept of frequency dynamic convolution, and the programmed algorithm in the official implementation code for frequency dynamic convolution takes slightly different procedure to reduce computational cost as in \cite{tdycnn}. While the programmed algorithm extracts frequency-adaptive attention weights in the same way, it applies convolution kernels differently. In the actual algorithm, outputs by each basis kernel are obtained first as in (1), then weighted sum is applied as in (2) as follows:

\begin{figure}[t]
\centerline{\includegraphics[width=8cm]{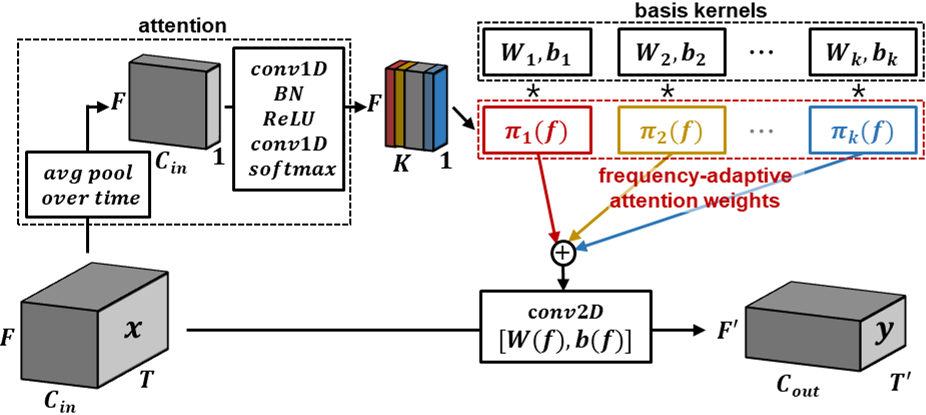}}
\vspace{-5pt}
\caption{An illustration of frequency dynamic convolution operation. $x$ and $y$ are input and output of frequency dynamic convolution layer. $T$, $F$ and $C_{in}$ are input dimension size of time, frequency and channel, and $T'$, $F'$ and $C_{out}$ are output dimension size of time, frequency and channel. $K$ is number of basis kernels, $W_i$ and $b_i$ are weight and bias of $i$th basis kernel and $\pi_i(f)$ is frequency-adaptive attention weight for $i$th basis kernel.}
\label{fig:fdyconv}
\vspace{-10pt}
\end{figure}

\begin{equation}
y_i(t,f) = W_i * x(t,f) + b_i
\end{equation}
\begin{equation}
y(t,f,x) = \sum_{i=1}^{K}\pi_i(f,x)y_i(t,f)
\end{equation}
\noindent
where $t$ is time, $f$ is frequency, $x$ and $y$ are input and output of one frequency dynamic convolution layer, $W_i$ and $b_i$ are weight and bias of $i$th basis kernel, $y_i$ is output from $i$th basis kernel and $\pi_i(f,x)$ is frequency-adaptive attention weight for $i$th basis kernel, and $K$ is number of basis kernels. This procedure is equivalent to the procedure illustrated in preceding paragraph with fewer computation.

%% file: sections/3_experimental_setups.tex
\subsection{Implementation Details}
\vspace{-5pt}
Input feature for SED model is a log Mel spectrogram with 128 Mel bins, extracted from 10-second-long audio data with sampling rate of 16kHz. SED models are trained with domestic environment sound event detection (DESED) dataset \cite{DCASEtask4} which consists of synthesized strongly labeled data, real weakly labeled data and real unlabeled data. Baseline is based on CRNN architecture \cite{crnn}. Attention pooling module is added at the last FC layer for joint training of weakly labeled data, and mean teacher method is applied for consistency training with unlabeled data for semi-supervised learning \cite{meanteacher}. Frame shift, mixup \cite{mixup}, time masking \cite{specaug} and FilterAugment \cite{filtaug} are applied for data augmentation. Baseline is the model using optimal step type FilterAugment from \cite{filtaug} with minor updates including seed of 21, mixup ratio of 1.0, and median filter that differs by classes. More details regarding the baseline are available in the GitHub repository and \cite{filtaug, DCASEtask4, dcasebaseline, mytechreport}.

\subsection{Evaluation Metrics}
\vspace{-5pt}
SED models were optimized to maximize sum of PSDS$_{1}$ and PSDS$_{2}$, which is the ranking score used in detection and classification of acoustic scenes and events (DCASE) 2021 and 2022 challenge task4 \cite{PSDS, dcasewebsite}. PSDS$_{1}$ favors SED system with accurate timestamps, while PSDS$_{2}$ favors SED system with less cross triggers. Collar-based F1 score and intersection-based F1 score with threshold of 0.5 are listed for the reference with labels “CB-F1” and “IB-F1” in the tables \cite{sedmetrics, insightdcase2020}.

Metric values listed in the tables are based on the best performance of each metric among 32 trained models, from student model and teacher model \cite{meanteacher} of 16 separate training runs. As dynamic convolutions are relatively unstable thus shows larger performance deviation, more training runs are operated compared to the previous work \cite{filtaug}. In addition, time taken to train models for 200 epochs using one NVIDIA RTX Titan are listed in Table 1 to compare training time for dynamic models.

%% file: sections/4_results_and_discussion.tex
\subsection{Dynamic Convolutions on SED}
\begin{table}[t]
\caption{Training time and SED performance of baseline, dynamic convolution \cite{dyconv}, temporal dynamic convolution \cite{tdycnn} and frequency dynamic convolution on DESED real validation dataset.}
\vspace{-5pt}
\centering
\setlength{\tabcolsep}{4.25pt}

\begin{tabular}{l|lllll}
\hline
\textbf{Models} & \textbf{Time} & \textbf{PSDS$_{1}$$\uparrow$} & \textbf{PSDS$_{2}$$\uparrow$} & \textbf{CB-F1$\uparrow$} & \textbf{IB-F1$\uparrow$} \\ \hline
baseline          & 3h 29m & 0.416          & 0.640          & 0.518          & 0.744          \\
DY \cite{dyconv}  & 6h 07m & 0.441          & 0.663          & 0.526          & 0.750          \\
TDY \cite{tdycnn} & 9h 23m & 0.415          & 0.652          & 0.512          & 0.751          \\
FDY               & 6h 10m & \textbf{0.452} & \textbf{0.670} & \textbf{0.533} & \textbf{0.753} \\ \hline
\end{tabular}
\vspace{-5pt}
\label{tab:dyconvs}
\end{table}

\vspace{-5pt}
We compared the performances of baseline with dynamic convolution \cite{dyconv}, temporal dynamic convolution \cite{tdycnn}, and proposed frequency dynamic convolution, abbreviated as DY-CRNN, TDY-CRNN and FDY-CRNN respectively. For all dynamic convolution models, dynamic convolution layers replaced all convolution layers except the first layer from the baseline model \cite{dyconv}, with four basis kernel and temperature of 31.

From the results in Table 1, FDY-CRNN significantly outperforms the rests. Dynamic convolution applies kernel that adapts to input as whole. Temporal dynamic convolution applies kernel that adapts to each time frame of input. Frequency dynamic convolution applies kernel that adapts to each frequency bin of input. Since kernel that adapts to each frequency bin of input significantly outperformed other content-adaptive kernels on SED, we can conclude that SED is a highly frequency-dependent task.

Considering that temporal dynamic convolution outperformed dynamic convolution on text-independent speaker verification by extracting speaker information from rapidly time-varying phonemes \cite{tdycnn}, it appears to be advantageous on SED as well because adapting kernels on time frames could help frame-wise predictions by SED. However, TDY-CRNN failed to outperform DY-CRNN and even marginally outperformed the baseline. It is because time-dependency is more critical on speaker verification due to the characteristics of speech data composed of short and rapidly changing phoneme sequences. Although time-dependency matters on SED too where sound events also vary along time axis, we should note that CRNN architecture for SED already process sequential information over time using recurrent neural network (RNN) layers. Therefore, TDY-CRNN is less effective than DY-CRNN is on SED, in terms of both performance and training time. On the contrary, FDY-CRNN that applies frequency-adaptive kernels performs much better because CRNN architecture lacks function to consider dependence on frequency regions.

\subsection{Number of Basis Kernels in FDY-CRNN}
\begin{table}[t]
\caption{FDY-SED performance with different number of basis kernels, $K$.}
\vspace{-5pt}
\centering
\setlength{\tabcolsep}{8.5pt}
\begin{tabular}{l|llll}
\hline
\textbf{$K$} & \textbf{PSDS$_{1}$ $\uparrow$} & \textbf{PSDS$_{2}$ $\uparrow$} & \textbf{CB-F1 $\uparrow$} & \textbf{IB-F1 $\uparrow$} \\ \hline
2 & 0.446          & 0.666          & 0.532          & 0.751          \\
3 & 0.449          & 0.668          & 0.531          & 0.754          \\
4 & \textbf{0.452} & \textbf{0.670} & 0.533          & 0.753          \\
5 & 0.445          & 0.665          & \textbf{0.540} & \textbf{0.755} \\ 
6 & 0.440          & 0.659          & 0.537          & 0.752          \\
\hline
\end{tabular}
\vspace{-10pt}
\label{tab:kernels}
\end{table}

\vspace{-5pt}
Number of basis kernels K directly affect the model’s representation capability and computational cost. Larger the K, stronger the expressiveness of trained SED model. However, too large K not only increase computational cost but also might cause overfitting of the model or undertraining of basis kernels. Table 2 shows SED performance of FDY-CRNN on different number of basis kernels. From the table, PSDS values are better on K=4 just as \cite{dyconv}, but F1 scores are better on K=5. As PSDS are more comprehensive metric that does not depend on calibrating threshold value while CB-F1 and IB-F1 depend on threshold value \cite{PSDS, sedmetrics, insightdcase2020}, we chose optimal model based on PSDS.

\subsection{Class-wise Performance Comparison between Baseline and FDY-CRNN}

\vspace{-5pt}
Class-wise performance of baseline and FDY-CRNN are compared for more detailed analysis on how frequency dynamic convolution affects SED performance. We chose representative models with following performances: baseline with PSDS$_1$ of 0.412, PSDS$_2$ of 0.634 and CB-F1 of 0.515 and FDY-CRNN with PSDS$_1$ of 0.432, PSDS$_2$ of 0.643 and CB-F1 of 0.532. Since PSDS is a comprehensive measure that takes into account of influences between different classes, class-wise performance is compared using CB-F1 instead.

It can be observed that baseline performs better on blender, frying, and vacuum cleaner in Figure 3. These quasi-stationary sound events are almost stationary over time \cite{randomdata}, thus exhibit simple time-frequency patterns as shown in Figure 4 (a), an example of vacuum cleaner sound log Mel spectrogram. Blender and vacuum cleaner sounds are mostly caused by running motors deriving dominant periodic mechanical noise \cite{vibration}. These sound events may involve other minor non-stationary noises such as blender cutting hard chunk or vacuum cleaner’s head hitting or rolling over other objects. Nonetheless, the motor sound is loud enough to dominate other noises thus these sound events can be considered quasi-stationary. Frying sound is caused by evaporation of water molecules on the surface of food being fried. Such evaporation occurs randomly and continuously just as raindrops randomly falling on ground, which is a classic example of random noise. Thus frying sound can be classified as random noise which is quasi-stationary as well \cite{randomdata}. Quasi-stationary sound events hardly change over time, thus result in horizontal patterns on log Mel spectrogram as shown in Figure 4 (a). Because such horizontal patterns are simple and similar on different frequency regions, the advantage of frequency dynamic convolution that it applies different kernels on different frequency regions is less evident for detecting quasi-stationary sound events.

\begin{figure}[t]
\centerline{\includegraphics[width=8cm]{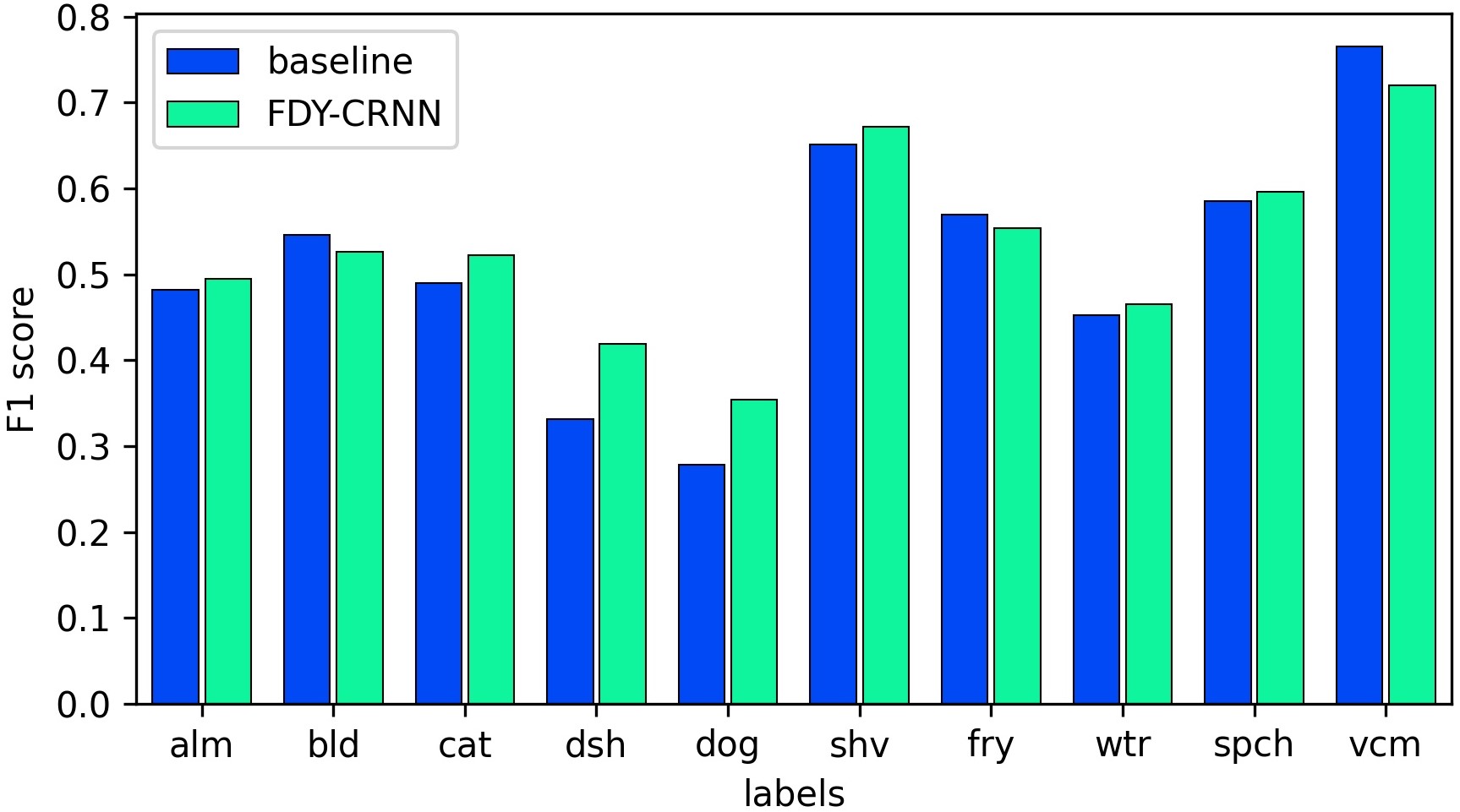}}
\vspace{-5pt}
\caption{Comparison of SED performance of baseline and FDY-CRNN in terms of class-wise collar-based F1 score. For labels, alm, bld, cat, dsh, dog, shv, fry, wtr, spch and vcm refer to alarm/bell ringing, blender, cat, dishes, dog, electric shaver/toothbrush, frying, running water, speech and vacuum cleaner respectively.}
\label{fig:classwisef1}
\vspace{-15pt}
\end{figure}

On the other hand, Figure 3 shows that FDY-CRNN performs better on the other sound event classes: alarm/bell ringing, cat, dishes, dog, electric shaver/toothbrush, running water and speech. These classes are non-stationary sound events those keep changing along time axis, thus result in intricate time-frequency patterns as shown in Figure 4 (b), an example of speech sound log Mel spectrogram. Alarm/bell ringing and dishes involve transient and abrupt short sounds. Cat, dog, and speech involve constantly changing pitches, with impulsive sounds such as stops and transient turbulent sound such as fricatives \cite{speechprocessing}. Electric shaver/toothbrush could be viewed as quasi-stationary sound like blender and vacuum cleaner because they are run by motors as well. However, their motor sound is not loud enough to dominate other impulsive noise they make while brushing teeth or shaving beard. Thus these sound events are rather non-stationary. Running water might appear as random noise like frying sound, as water that keeps running alone involves turbulent sound as it hits other surfaces \cite{randomdata}. But running water in domestic environments involves interactions with people, as people would not just let it flow for no reason in their home. Human interaction keeps intervening the sound of running water, thus it is considered as a non-stationary sound event. Non-stationary sound events keeps changing its frequency components over time, resulting in more intricate patterns on various frequency regions of log Mel spectrogram as shown in Figure 4 (b).

\begin{figure}[t]
\centerline{\includegraphics[width=8.25cm]{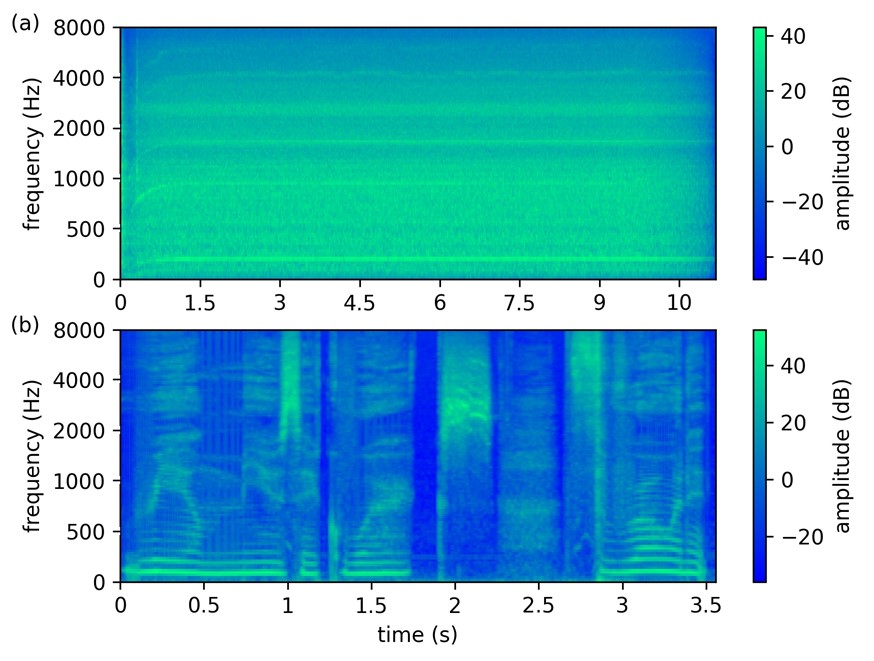}}
\vspace{-10pt}
\caption{Log Mel spectrogram examples of quasi-stationary and non-stationary sound events: (a) vacuum cleaner sound event as an example of quasi-stationary sound, (b) speech sound event as an example of non-stationary sound.}
\label{fig:vacuumspeech}
\vspace{-15pt}
\end{figure}

From above discussions, it could be inferred that frequency dynamic convolution has greatly improved SED performance by enhancing recognition of diverse and intricate patterns that non-stationary sound events exhibit, by applying frequency-adaptive kernels. This result again proves the premise on this work that frequency dynamic convolution effectively recognizes frequency-dependent patterns of sound events.

%% file: sections/5_conclusion.tex
Frequency dynamic convolution is proposed to recognize frequency-dependent patterns of sound event data for SED. Conventional 2D convolution imposes translation equivariance along both time and frequency axis, but it is physically inconsistent with frequency-dependent patterns of sound events. Thus frequency dynamic convolution is designed to release translation equivariance along frequency axis by applying frequency-adaptive kernels and enforce physical consistency of the model with time-frequency patterns in sound events. Experiments on DESED dataset showed that frequency dynamic convolution is superior to not only baseline but also dynamic convolution and temporal dynamic convolution. In addition, comparison of class-wise F1 scores between baseline and FDY-CRNN showed that frequency dynamic convolution is especially helpful in detection of non-stationary sound events, proving effectivity of frequency dynamic convolution on frequency-dependent patterns.

%% file: paper.bbl
\begin{thebibliography}{10}
\providecommand{\url}[1]{#1}
\csname url@samestyle\endcsname
\providecommand{\newblock}{\relax}
\providecommand{\bibinfo}[2]{#2}
\providecommand{\BIBentrySTDinterwordspacing}{\spaceskip=0pt\relax}
\providecommand{\BIBentryALTinterwordstretchfactor}{4}
\providecommand{\BIBentryALTinterwordspacing}{\spaceskip=\fontdimen2\font plus
\BIBentryALTinterwordstretchfactor\fontdimen3\font minus
  \fontdimen4\font\relax}
\providecommand{\BIBforeignlanguage}[2]{{%
\expandafter\ifx\csname l@#1\endcsname\relax
\typeout{** WARNING: IEEEtran.bst: No hyphenation pattern has been}%
\typeout{** loaded for the language `#1'. Using the pattern for}%
\typeout{** the default language instead.}%
\else
\language=\csname l@#1\endcsname
\fi
#2}}
\providecommand{\BIBdecl}{\relax}
\BIBdecl

\bibitem{CASSE}
T.~Virtanen, M.~D. Plumbley, and D.~Ellis, \emph{Computational Analysis of
  Sound Scenes and Events}, 1st~ed.\hskip 1em plus 0.5em minus 0.4em\relax
  Springer Publishing Company, Incorporated, 2017, pp. 3--11, 71--77.

\bibitem{sedmetrics}
A.~Mesaros, T.~Heittola, and T.~Virtanen, ``Metrics for polyphonic sound event
  detection,'' \emph{Applied Sciences}, vol.~6, no.~6, 2016.

\bibitem{DCASEtask4}
N.~Turpault, R.~Serizel, A.~Parag~Shah, and J.~Salamon, ``{Sound event
  detection in domestic environments with weakly labeled data and soundscape
  synthesis},'' in \emph{{Workshop on Detection and Classification of Acoustic
  Scenes and Events}}, 2019.

\bibitem{mytechreport}
H.~Nam, B.-Y. Ko, G.-T. Lee, S.-H. Kim, W.-H. Jung, S.-M. Choi, and Y.-H. Park,
  ``Heavily augmented sound event detection utilizing weak predictions,''
  DCASE2021 Challenge, Tech. Rep., 2021.

\bibitem{dcasewebsite}
\BIBentryALTinterwordspacing
DCASE. Dcase 2021 challenge task4: Sound event detection and separation in
  domestic environmentse. [Online]. Available:
  \url{http://dcase.community/challenge2021/task-sound-event-detection-and-separation-in-domestic-environments}
\BIBentrySTDinterwordspacing

\bibitem{specaug}
D.~S. Park, W.~Chan, Y.~Zhang, C.-C. Chiu, B.~Zoph, E.~D. Cubuk, and Q.~V. Le,
  ``{SpecAugment: A Simple Data Augmentation Method for Automatic Speech
  Recognition},'' in \emph{Proc. Interspeech}, 2019, pp. 2613--2617.

\bibitem{conformer}
A.~Gulati, J.~Qin, C.-C. Chiu, N.~Parmar, Y.~Zhang, J.~Yu, W.~Han, S.~Wang,
  Z.~Zhang, Y.~Wu, and R.~Pang, ``{Conformer: Convolution-augmented Transformer
  for Speech Recognition},'' in \emph{Proc. Interspeech}, 2020, pp. 5036--5040.

\bibitem{coughcam}
G.-T. Lee, H.~Nam, S.-H. Kim, S.-M. Choi, Y.~Kim, and Y.-H. Park, ``Deep
  learning based cough detection camera using enhanced features,'' \emph{Expert
  Systems with Applications}, vol. 206, 2022.

\bibitem{acnn}
S.-H. Kim and Y.-H. Park, ``Adaptive convolutional neural network for
  text-independent speaker recognition,'' in \emph{Proc. Interspeech}, 2021,
  pp. 641--645.

\bibitem{transformer}
A.~Vaswani, N.~Shazeer, N.~Parmar, J.~Uszkoreit, L.~Jones, A.~N. Gomez,
  L.~Kaiser, and I.~Polosukhin, ``Attention is all you need,'' in
  \emph{Advances in Neural Information Processing Systems}, 2017.

\bibitem{bert}
J.~Devlin, M.-W. Chang, K.~Lee, and K.~Toutanova, ``{BERT}: Pre-training of
  deep bidirectional transformers for language understanding,'' in
  \emph{Proceedings of the 2019 Conference of the North {A}merican Chapter of
  the Association for Computational Linguistics}, 2019, pp. 4171--4186.

\bibitem{dcase7wang}
Y.-W. Wang, C.-P. Chen, C.-L. Lu, and B.-C. Chan, ``{CHT+NSYSU Sound Event
  Detection System With Multiscale Channel Attention And Multiple Consistency
  Training For DCASE 2021 Task 4},'' DCASE2021 Challenge, Tech. Rep., 2021.

\bibitem{dcase22koo}
H.~Koo, H.-M. Park, J.~Park, and M.~Oh, ``Sound event detection based on
  self-supervised learning of wav2vec 2.0,'' DCASE2021 Challenge, Tech. Rep.,
  2021.

\bibitem{dcase18hang}
Y.~Chen, ``Convolution-augmented conformer for sound event detection,''
  DCASE2021 Challenge, Tech. Rep., 2021.

\bibitem{dcase21na}
T.~Na and Q.~Zhang, ``Convolutional network with conformer for semi-supervised
  sound event detection,'' DCASE2021 Challenge, Tech. Rep., 2021.

\bibitem{dcase3}
R.~Lu, W.~Hu, D.~Zhiyao, and J.~Liu, ``Integrating advantages of recurrent and
  transformer structures for sound event detection in multiple scenarios,''
  DCASE2021 Challenge, Tech. Rep., 2021.

\bibitem{tdycnn}
S.-H. Kim, H.~Nam, and Y.-H. Park, ``Temporal dynamic convolutional neural
  network for text-independent speaker verification and phonemetic analysis,''
  \emph{International Conference on Acoustics, Speech and Signal Processing
  (ICASSP)}, 2022.

\bibitem{DTDY}
S.~H. Kim, H.~Nam, and Y.~H. Park, ``Decomposed temporal dynamic cnn: Efficient
  time-adaptive network for text-independent speaker verification explained
  with speaker activation map,'' \emph{arXiv preprint arXiv:2203.15277}, 2022.

\bibitem{filtaug}
H.~Nam, S.-H. Kim, and Y.-H. Park, ``Filteraugment: An acoustic environmental
  data augmentation method,'' \emph{International Conference on Acoustics,
  Speech and Signal Processing (ICASSP)}, 2022.

\bibitem{dyconv}
Y.~Chen, X.~Dai, M.~Liu, D.~Chen, L.~Yuan, and Z.~Liu, ``Dynamic convolution:
  Attention over convolution kernels,'' in \emph{Proceedings of the IEEE/CVF
  Conference on Computer Vision and Pattern Recognition}, June 2020.

\bibitem{PSDS}
{\c{C}}.~Bilen, G.~Ferroni, F.~Tuveri, J.~Azcarreta, and S.~Krstulovi\'{c}, ``A
  framework for the robust evaluation of sound event detection,'' in
  \emph{International Conference on Acoustics, Speech and Signal Processing
  (ICASSP)}, 2020, pp. 61--65.

\bibitem{crnn}
E.~{\c{C}}ak{\i}r, G.~Parascandolo, T.~Heittola, H.~Huttunen, and T.~Virtanen,
  ``Convolutional recurrent neural networks for polyphonic sound event
  detection,'' \emph{IEEE/ACM Transactions on Audio, Speech, and Language
  Processing}, vol.~25, no.~6, pp. 1291--1303, 2017.

\bibitem{physicsinformedML}
G.~E. {Karniadakis}, I.~G. {Kevrekidis}, L.~{Lu}, P.~{Perdikaris}, S.~{Wang},
  and L.~{Yang}, ``{Physics-informed machine learning},'' \emph{Nature Reviews
  Physics}, vol.~3, no.~6, pp. 422--440, Jan. 2021.

\bibitem{vit}
A.~Dosovitskiy, L.~Beyer, A.~Kolesnikov, D.~Weissenborn, X.~Zhai,
  T.~Unterthiner, M.~Dehghani, M.~Minderer, G.~Heigold, S.~Gelly, J.~Uszkoreit,
  and N.~Houlsby, ``An image is worth 16x16 words: Transformers for image
  recognition at scale,'' in \emph{9th International Conference on Learning
  Representations, {ICLR} 2021, Virtual Event, Austria, May 3-7, 2021}, 2021.

\bibitem{meanteacher}
A.~Tarvainen and H.~Valpola, ``Mean teachers are better role models:
  Weight-averaged consistency targets improve semi-supervised deep learning
  results,'' in \emph{Advances in Neural Information Processing Systems},
  vol.~30, 2017.

\bibitem{mixup}
H.~Zhang, M.~Cisse, Y.~N. Dauphin, and D.~Lopez-Paz, ``mixup: Beyond empirical
  risk minimization,'' in \emph{International Conference on Learning
  Representations}, 2018.

\bibitem{dcasebaseline}
\BIBentryALTinterwordspacing
N.~Turpault. Dcase2021 task4 baseline. GitHub. Available:
  https://github.com/DCASE-REPO/DESED\_task. [Online]. Available:
  \url{https://github.com/DCASE-REPO/DESED\_task}
\BIBentrySTDinterwordspacing

\bibitem{insightdcase2020}
G.~Ferroni, N.~Turpault, J.~Azcarreta, F.~Tuveri, R.~Serizel,
  {\c{C}a\v{g}da\c{s}}.~Bilen, and S.~Krstulovi\'{c}, ``Improving sound event
  detection metrics: Insights from dcase 2020,'' in \emph{International
  Conference on Acoustics, Speech and Signal Processing (ICASSP)}, 2021, pp.
  631--635.

\bibitem{randomdata}
J.~Bendat and A.~Piersol, \emph{Random Data: Analysis and Measurement
  Procedures}, 4th~ed.\hskip 1em plus 0.5em minus 0.4em\relax Wiley, 2011, pp.
  8--12, 123.

\bibitem{vibration}
D.~Inman, \emph{Engineering Vibrations}, 4th~ed.\hskip 1em plus 0.5em minus
  0.4em\relax Pearson, 2013, pp. 172--177.

\bibitem{speechprocessing}
L.~Rabiner and R.~Schafer, \emph{Theory and Applications of Digital Speech
  Processing}, 1st~ed.\hskip 1em plus 0.5em minus 0.4em\relax Pearson, 2010,
  pp. 89--123.

\end{thebibliography}
